\documentclass[letterpaper]{article}
\usepackage{aaai17}
\usepackage{times}
\usepackage{helvet}
\usepackage{courier}
\usepackage{graphicx}
\usepackage{url}

\usepackage{color}
\usepackage{tikz}
\usepackage{subfigure}
\usetikzlibrary{mindmap,trees,backgrounds}

\usetikzlibrary{arrows,decorations.pathmorphing,backgrounds,fit,positioning,shapes.symbols,chains,shapes}
\newcommand{\myunit}{0.1 cm}
\tikzset{node style sp/.style={draw,circle,minimum size=\myunit}}
\tikzset{node style el/.style={draw,ellipse,minimum size=\myunit}}

\title{Argumentation-based Security for Social Good}

\author{Erisa Karafili$^*$, Antonis C. Kakas$^+$,  Nikolaos I. Spanoudakis$^\dagger$, and Emil C. Lupu$^*$\\ 
$^*$Department of Computing, Imperial College London\\
$^+$Department of Computer Science, University of Cyprus\\
$^\dagger$Applied Mathematics and Computers Lab., Technical University of Crete\\
e.karafili@imperial.ac.uk, antonis@ucy.ac.cy, nispanoudakis@isc.tuc.gr,  e.c.lupu@imperial.ac.uk
}

\begin{document}
\maketitle
\begin{abstract}
The increase of connectivity and the impact it has in every day life is raising new and existing security problems 
that are becoming important for social good. 
We introduce two particular problems: cyber attack attribution and regulatory data sharing.
For both problems, decisions about which rules to apply, should be taken
under incomplete and context dependent information. 
The solution we propose is based on argumentation reasoning, that is
a well suited technique for implementing decision making mechanisms under conflicting and incomplete information.
Our proposal permits us to identify the attacker of a cyber attack
and decide the regulation rule that should be used while using and sharing data. 
 We illustrate our solution through concrete examples.
\end{abstract}

\section{Introduction}
The interconnectivity and the  increased use of ``smart'' connected devices 
is improving our daily life, e.g., smart homes, e-health, elder care, AI surgical robots, smart cars.
The growing impact of these devices 
 is transforming them into targets of cyber attacks, e.g., remote attacks made to cars' autopilot, 
private information leak due to cyber attacks in cloud, attacks made on medical devices.
Protecting the privacy of these devices as well as the information used by them is still an important problem,
and becomes a pressing one in our highly interconnected cyber world.
One of the challenges that this problem raises is ensuring the usability of the system
while protecting it. 

An effective way for protecting the system from cyber attacks is to enforce specific protective measures
that are specific to the attacker performing the attack. Hence, identifying the attacker makes the protective actions more adequate 
and efficient. 
Discovering the attacker is not a trivial task, as the information that we are
working with is incomplete, context dependent and the attacker use anti-forensic tools to hide themselves.

Connectivity has increased the amount of shared and used data, but also underlined 
the need to protect and correctly use and access the data.
Also in this case, different knowledge should be taken into consideration before deciding the regulation rules
to be applied. For data sharing the knowledge is well documented in 
regulation and/or legislative rules, but dealing with them,
especially when different rules can be applied simultaneously, is not an easy task. 

Deciding which protective measures to take, what information to use while identifying the attacker,
or what rule/law to apply for a particular piece of data used by a specific user are all decision problems. 
The decision is made under a certain grade of uncertainty as the information is not complete,
it can be contradictory and context dependent. 
We decided to represent and solve the above problems by using argumentation-based reasoning.
Argumentation-based reasoning, for simplicity \emph{argumentation reasoning}, is a well suited technique for implementing decision making mechanisms under conflicting,
incomplete and context related knowledge. 

We believe argumentation reasoning is an appropriate technique for solving the above cyber security problems, as 
we have successfully applied it to rule prioritization in other security areas~\cite{BandaraKLR09}. 
 Furthermore,  a first prototype system for data access of medical data according to 
european union and national law and based on argumentation
 had been developed\footnote{\url{http://www.cs.ucy.ac.cy/projects/health-record/login_form.html}}.

We construct a reasoning process for the cyber attack problem that decides which rule and information to use,
depending on the given information for identifying the attacker. 
For the data sharing problem we construct a decision process model that given the different regulation and legislative
rules, decides depending on the context how and who can access the data and the rules that should be applied.
\vspace{-10pt}
\paragraph{Argumentation background}
We review briefly the basic theory of argumentation which we will use to model our security problems.
The theory will be presented from a general point of view of applying argumentation to real-life (decision) problems.

In~\cite{KMD94,gorgias1} a preference-based argumentation framework was proposed for representing 
multi-agent application problems via argumentation composed of different levels. 
\emph{Object level arguments} represent the possible decisions or actions in a specific application domain and 
\emph{first-level priority arguments} express preferences on the object level arguments in order to resolve possible conflicts. 
\emph{Higher-order priority arguments} are used to resolve conflicts between priority arguments of the previous level.
For sake of simplicity, we call \emph{arguments rule} the objective level arguments, and \emph{priority rules}
both types of priority arguments.

An \emph{argumentation theory} is a pair ($\cal T,\cal P$) composed of formulae 
 of the form $L \leftarrow L_1, \ldots,L_n$, where $L,L_1,\ldots,L_n$ are positive or negative ground literals. 
 $\cal T$ are the arguments rule, $\cal P$ are the priority rules. For $\cal P$ the head $L$ has the form
 $L= rule_1 > rule_2$, where $rule_1$ and $rule_2$ are two rules, while $>$ refers to an (irreflexive) higher priority
 relation amongst them, it means $rule_1$ has higher priority than $rule_2$.
 The arguments are given by subsets $(T, P)$, where $T \subseteq\cal T$ and $P \subseteq \cal P$.
 An argument $(T, P)$ supports its conclusion $L$, when $L$ is derived by $T$ and supported by $P$.
 A counter-argument for $(T, P)$ is another argument $(T', P')$, where they derive two contrary conclusions (i.e., $L$ and $\neg L$).
 For sake of simplicity, we call the pair of argument and counter-argument, \emph{conflicting} arguments.
 
Decisions are given by the admissible arguments of the given argumentation theory extended to the application scenario.
An argument is admissible if its priority rules are stronger than its counter-arguments. 

Argumentation is well suited for solving problems under   \emph{partial information} 
 as only the arguments whose premisses are satisfied will play a role. 
 It is important to address the \emph{reverse engineering problem} of finding additional conditions 
 under which a certain option will be a solution. This is achieved by 
  combining abductive reasoning with argumentation to find strong arguments under the
   assumption that some further abducible assumptions hold in a scenario.
Abductive reasoning combined with argumentation are the core of the GorgiasB tool\footnote{\url{http://gorgiasb.tuc.gr/}}~\cite{gorgias2} which permits
to develop applications under preference-based argumentation. 

\section{The Attribution Problem in Cyber Attacks}
 The increase of connectivity together with the expansion of Internet of Things (IoT) can
lead to a drastic increase of cyber attacks. 
 Protective measures on their own are not enough, because for being effective they need to be specifically constructed 
depending on the types of attack and
attacker. 
Forensics investigation can determine the type of attack and support the discovery of the attacker.
The latter is not trivial due to anti-forensics tools used by the attacker and the dynamicity of attacks.
We call \emph{attribution problem} the problem of finding out who the attacker is.
Forensics in cyber attacks helps during the attribution problem by collecting the evidence. 
The ability of finding through forensics investigation who the attacker was, makes the new attacks more expensive, 
as the attackers have to hide their identities. 

\paragraph{Importance of attribution in social good}
The growing impact that connectivity 
and connected devices have  
in every day life, e.g., smart homes, elder care, e-health, 
surveillance and security,  is making them more sensitive to attacks. 
 Attribution can relieve part of these attacks, by reducing and mitigating them.
Knowing the attacker permits to enforce specific attacker oriented counter measures,  
reduces the effects of the attacks, and diminishes the occurrence 
of future attacks performed by the same attacker. 
Finding who performed the attack not only helps in  protecting the system, but also for 
bringing perpetrators to justice.

The attribution problem is important when dealing with multilateral relations between countries/companies/entities. 
Nowadays, cyber attacks
to specific government institutions, companies, or entities by groups of hackers that are coming from (or working for)
a competitor/rival group/country, or group of terrorists, are escalating. The attacks that are not or badly attributed can 
aggravate diplomatic 
relations between countries. Finding out swiftly who the original sender and/or executer of a cyber attack was can help also in avoiding 
diplomatic mistakes between countries or entities.

 Attribution and forensics come along with their own challenges: difficulties to collect evidence, 
 its high costs, and the vastness of anti-forensics tools used by the attackers. 
 In social sciences (e.g., war science~\cite{rid}, criminology~\cite{Hober}, sociology~\cite{social}) some attribution techniques have been developed, but 
 currently there is no implementation of these techniques in forensics for cyber attacks.
   Attribution has to constantly cope with incomplete and conflicting evidence.
   \vspace{-10pt}
  \paragraph{Main idea of our approach} We introduce a reasoner that given pieces of evidence,  
  it deduces wether a given country performed or not the attack. 
  We first build up and categorize the evidence, by using a social science technique, called the Q-Model~\cite{rid}. 
  This model permits us to reason and categorize the evidence collected during the forensics investigation, 
  which is going to be used by our reasoner for attributing the cyber attacks.
  The evidence is used as premisses for the reasoning rules that permit us to reach a conclusion about the origin 
  of the attack.
  The reasoning rules are extracted during the analysis of different attribution forensics examples. 
  
  The collected and categorized evidence can be incomplete and conflicting. 
 We decided to use argumentation and abductive reasoning to  represent the different evidence and reasoning rules,
 and to construct our reasoner. 
This approach is able to cope with incomplete information through the abductive reasoning, which permits
to get good plausible explanations for a given piece
 of evidence. 
 On the other hand, argumentation reasoning permits to work with conflicting information. 
 We model the reasoning rules as arguments and refer to conflicting evidence as conflicting arguments, i.e.,
 evidence used by reasoning rules that derive contrary conclusions.
 Argumentation reasoning deals with conflicting evidence by putting preferences between arguments, through priority rules.
    The argumentation approach provides an explanation to the solutions/decisions that it makes. This helps the analyst/user to
accept or reject the recommendations of the reasoner and, in general, to trust its use.
To the best of our knowledge, this is the first attempt to use a social science technique for attribution
together with argumentation for solving the attribution problem in cyber attacks.
   To construct our reasoner we use the GorgiasB tool.

 \subsection{Argumentation-based Reasoner for Attribution}
 Let us introduce our reasoner for the attribution problem in cyber attacks.
 We show how we can categorize the different pieces of evidence. 
 We introduce the reasoning rules extracted from a simple cyber attack example and transformed them into arguments.
 Finally, we show the use of arguments and priority rules for solving the attribution problem.

 \vspace{-10pt}
\paragraph{Collecting and categorizing evidence}
 \begin{figure}[htb]
\centering
\scalebox{0.8}{
 \begin{tikzpicture}[node distance = 1cm, auto,font=\scriptsize,
axis/.style={very thick, ->, >=stealth'},
every node/.style={node distance=3cm},
comment/.style={font=\scriptsize\sffamily},
c1/.style={draw,circle, inner sep=0.1cm, text badly centered,  font=\scriptsize\sffamily},
c2/.style={draw,circle, inner sep=1.3cm, text badly centered, font=\scriptsize\sffamily},
c3/.style={draw,circle, inner sep=1.9cm,text badly centered, font=\scriptsize\sffamily},
b/.style = {circle,inner sep=0pt,fill=black,minimum size=1mm,draw=black}] 

\node[c1](circ1){\textbf{STRATEGIC}};
\node[c2](circ2){};
\node[c3](circ3){};

\node[above=0.2 of circ1](t1){\textbf{OPERATIONAL}};
\node[above=1.2 of circ1](t2){\textbf{TACTICAL}};

\node[right=0.4 of circ1](i21){};

\node[below=0.15 of circ1](t21){\emph{capability}};
\node[above=0.07 of i21](t22){\emph{motive}};
\node[below=0.4 of t22](t24){\emph{damage}};

\node[left=1.7 of t22](t23){\emph{context}};
\node[below=0.4 of t23](t25){\ \emph{claims}};

\node[left=0.2 of t22](t11){\emph{why}};
\node[left=0.9 of t22](t13){\emph{who}};

\node[above=0.25 of t21](t12){\emph{objectives}};

\node[below=0.6 of t21](t31){\emph{domain name}};

\node[below=0.8 of t24](n32){ };

\node[left=2.2 of n32](t33){\emph{language}};
\node[right=2.05 of t33](t32){\emph{personas}};

\node[left=1 of circ1](t33){\emph{target}};

\node[right=1.2 of circ1](t34){\emph{IP}};

\node[left=0.6 of t1](t35){\emph{avoid}};
\node[right=0.5 of t1](t35){\emph{code}};
\end{tikzpicture}}
  \vspace{-8pt}
 \caption{The Q-Model with the different layers \label{fig:qmodel}}
\end{figure}
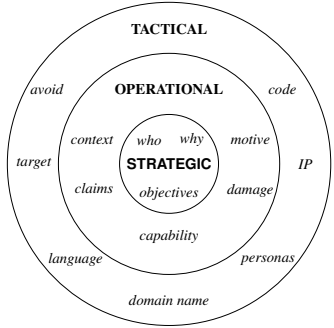

For collecting and categorizing the evidence we use the Q-Model.
  The latter  helps the analysts asking a full range of possible questions and considering all possible evidence, arguments and conclusions.
 The Q-Model divides evidence in three layers as shown in Figure~\ref{fig:qmodel}. The \emph{tactical layer} includes technical 
 information, obtained  by checking the system, its logs, and the attack code:
 IP addresses from where the attack came from, domain names, the original language of the programmers that
 wrote the code, the target in the system.
  The \emph{operational layer} includes less technical evidences, e.g., 
 entity or group claiming the attack, the context (e.g., political or geographical) of the attack, the capability/skills required for performing the attack.
 The \emph{strategic layer} includes high level evidence collected by 
 analyzing 
 the attacked entity and interviewing strategic people inside it,
  e.g., how successful the attack was and what objectives it had, who performed the attack. 
  Given an attack the pieces of evidence are collected by answering the questions of each layer,
 and categorized.

 \paragraph{Constructing the arguments}
 Given the evidence, we need  to extract from existing forensics analyses, the various reasoning rules and represent 
 them through argumentation reasoning, that is composed of the different arguments (arguments rules and their priorities) and their
 conclusions. 
 We construct from the reasoning rules the arguments rules and their priority rules,
 where the collected evidence are the premisses that support an argument for reaching a certain conclusion,
 while the conclusion is the statement about
a given country, if it is the attacker or not.

 Let's see an example. Suppose that a network is under a \emph{ssh attack}. The analyst, as s/he has no evidence yet, 
 cannot say if a particular country performed the attack.
  \begin{itemize}
 \item If no evidence is given for a particular country, then it did not performed the attack.
  \end{itemize}
  The above rule is represented as an argument rule below.
 \begin{equation}\label{1rule} 
\textrm{not }  \textrm{perform}(\textrm{Attack}, \textrm{Country}) \leftarrow 
\end{equation}
The analyst starts by checking the system logs, and finds a couple of IPs that were involved in the attack. Part of the IPs
 are found to come from country $C1$, thus, the conclusion is that $C1$ performed the attack, as in rule~\ref{ip}.
  \begin{itemize}
 \item If the attack came from IP1, and IP1 is in country C1, then C1 performed the attack.
 \end{itemize}
  \begin{equation}\label{ip}
  \begin{array}{ll}
\textrm{perform}(\textrm{Attack}, \textrm{Country})  \leftarrow & \textrm{sourceIP}(\textrm{Attack}, IP) \ \wedge\\ 
&  \textrm{geoloc}(IP, \textrm{Country})
\end{array}
\end{equation}
Further analyses revealed that the above $IPs$ were spoofed (anonymized), hence they cannot be used to link with $C1$. Therefor, 
$C1$ did not performed the attack. 
 \begin{itemize}
 \item If the attack came from IP1, and IP1 is in C1, but
IP1 is spoofed, then C1 did not performed the attack.
\end{itemize}
  \begin{equation}\label{spoofed} 
    \begin{array}{r}
\textrm{not } \textrm{perform}(\textrm{Attack}, \textrm{Country}) \leftarrow \textrm{sourceIP}(\textrm{Attack}, IP)\wedge\\ \textrm{geoloc}(IP, \textrm{Country}) \wedge \textrm{spoofed}(IP)
\end{array}
\end{equation}
Rule~\ref{spoofed} is in conflict with rule~\ref{ip}, as they are concluding opposite statements. In this case,
 rule~\ref{spoofed} is stronger then rule~\ref{ip}, because spoofed(IP) is a stronger argument then the geolocation of the IP. 
  Thus, their priority rule is rule~\ref{spoofed} $>$ rule~\ref{ip}.

The analyst, by performing a reverse engineering of the attacks code, finds that the attack was build for avoiding $C1$. 
\begin{itemize}
  \item If  the attack was designed for avoiding a specific country C1, then C1 performed the attack. 
 \end{itemize}
\begin{equation}\label{avoid} 
      \begin{array}{r}
\textrm{perform}(\textrm{Attack}, \textrm{Country}) \leftarrow \textrm{avoid}(\textrm{Attack}, \textrm{Country})
\end{array}
\end{equation}
When we take into account the rest of the evidence, still $C1$ performed the attack
 as avoid is a stronger argument (rule~\ref{avoidp}).
\begin{itemize}
  \item If the attack was designed for avoiding a specific country, C1, even if
  the IPs that link the attack to country C1 are spoofed, C1 performed the attack. 
 \end{itemize}
\begin{equation}\label{avoidp} 
      \begin{array}{r}
\textrm{perform}(\textrm{Attack}, \textrm{Country}) \leftarrow \textrm{avoid}(\textrm{Attack}, \textrm{Country}) \ \wedge \\
 \textrm{sourceIP}(\textrm{Attack}, IP) \wedge\textrm{geoloc}(IP, \textrm{Country})  \wedge\\ \textrm{spoofed}(IP) 
\end{array}
\end{equation}
In this case, rule~\ref{avoidp} and \ref{spoofed} are conflicting arguments, as they conclude two opposite statements.
 Rule~\ref{avoidp} is stronger than rule~\ref{spoofed} (rule~\ref{avoidp} $>$ rule~\ref{spoofed}) as avoid is a stronger evidence.

\begin{figure}[!htb]
\centering
\scalebox{0.75}{
\begin{tikzpicture}[node distance = 1cm, auto,font=\scriptsize,
axis/.style={very thick, ->, >=stealth'},
every node/.style={node distance=3cm},
comment/.style={font=\scriptsize\sffamily},
pred/.style={rectangle, draw, inner sep=0.2cm, text width=4.5cm, text badly centered, font=\scriptsize\sffamily},
pred1/.style={rectangle, draw, inner sep=0.2cm, text width=1.8cm, text badly centered, font=\scriptsize\sffamily},
b/.style = {circle,inner sep=0pt,fill=black,minimum size=1mm,draw=black}] 

\node [pred](a1) {$A1:\textrm{not } \textrm{perform}(\textrm{Attack}, \textrm{Country})$};

\node [pred, below=0.2 of a1](a2) {$A2:\textrm{perform}(\textrm{Attack}, \textrm{Country})$ When
 $\textrm{sourceIP}(\textrm{Attack}, IP), \textrm{geoloc}(IP, \textrm{Country})$};

\node [pred, below=0.2 of a2](a3) {$A3:\textrm{perform}(\textrm{Attack}, \textrm{Country})$ When
 $\textrm{sourceIP}(\textrm{Attack}, IP), \textrm{geoloc}(IP, \textrm{Country})$, $\textrm{spoofed}(IP)$};
 
 \node [pred, below=0.2 of a3](a4) {$A4:\textrm{perform}(\textrm{Attack}, \textrm{Country})$ When
 $\textrm{avoid}(\textrm{Attack}, \textrm{Country})$};
 
  \node [pred, below=0.2 of a4](a5) {$A5:\textrm{perform}(\textrm{Attack}, \textrm{Country})$ When
 $\textrm{avoid}(\textrm{Attack}, \textrm{Country}), \textrm{spoofed}(IP),$  $\textrm{sourceIP}(\textrm{Attack}, IP), \textrm{geoloc}(IP, \textrm{Country})$};

\node [pred1, right=1 of a1)](b1) {$B1: A2 > A1$};
\node [pred1, below=1 of b1)](b2) {$B2: A1 > A3$};
\node [pred1, below=1 of b2)](b3) {$B3: A4 > A1$};
\node [pred1, below=1 of b3)](b4) {$B4: A5 > A1$};

\node [left=0.984 of b1](pp1) {};
\node [below=1 of pp1](ppp1) {};
\node [above=1.128 of ppp1](p1) {};

\node [left=0.93 of b1](pp3) {};
\node [below=1 of pp3](ppp3) {};
\node [above=1.128 of ppp3](p13) {};

\node [left=0.91 of b1](pp4) {};
\node [below=1 of pp4](ppp4) {};
\node [above=1.128 of ppp4](p14) {};

\node [left=1 of b2](ab) {};
\node [above=0.24 of ab](ab2) {};

\node [left=1 of b3](p3) {};

\node [right=0.91 of a3](pb3) {};
\node [below=0.5 of pb3](pb31) {};

\node [right=1 of a2](bb2) {};
\node [below=0.15 of bb2](pb2) {};

\node [right=0.9 of a5](pb4) {};

\node [above=0.1 of pb4](pb41) {};

\node [above=1 of pb2](ab1) {};

\path[ ->] (a1) edge node {} (b1);
\path[ ->] (ab2) edge node {} (ab1);

\path[ ->] (p1) edge node {} (pb2);
\path[ ->] (a3) edge node {} (b2);

\node [right=0.93 of a4](ab4) {};
\node [below=0.5 of ab4](bb4) {};

\node [above=1.28 of bb4](bb3) {};

\path[ ->] (p13) edge node {} (bb3);
\path[ ->] (a4) edge node {} (b3);

\node [right=0.91 of a4](ab4) {};
\node [below=0.58 of ab4](bb4) {};

\path[ ->] (p14) edge node {} (bb4);
\path[ ->] (a5) edge node {} (b4);

\node [pred1, right=4 of a3)](c1) {$C1: B4 > B2$};
\node [right = 4 of p3](c11) { };
\node [above = 0.36 of c11](c12) { };

\path[ ->] (b2) edge node {} (c1);
\path[ ->] (b4) edge node {} (c12);

\end{tikzpicture}}
\vspace{-20pt}
\caption{Decision diagram showing the reasoning rules \label{fig:example1}}
\vspace{-15pt}
\end{figure}
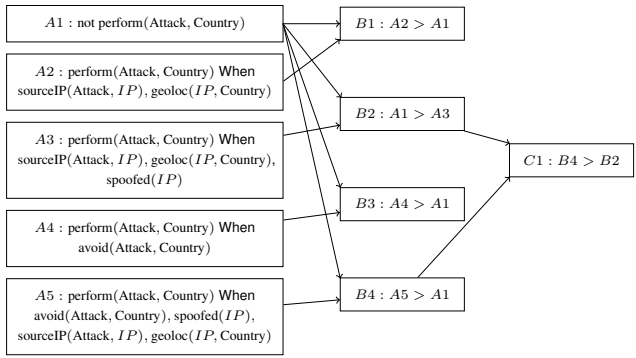

 \paragraph{Attribution with the reasoning rules}
The above rules are given to GorgiasB, which allows us to define the arguments, to capture and resolve their conflicts by introducing
the various priority rules. 
 In Figure~\ref{fig:example1}, we show how the above rules can be used 
 for reaching the conclusion, and the preferences that are used.
 We assign to every rule a label, e.g., $A1$, $B2$.  
 For example, the priority rule $B1$ represents that $A2$ is stronger than $A1$ ($A2 >A1$),
where $A1$ represents rule~\ref{1rule} and $A2$ rule~\ref{ip}.
    
    An example involving conflicting arguments can be traced between rules B2 and B4, and the priority
    between them is given by C1. B2 represents rule~\ref{spoofed}, while B4 represents rule~\ref{avoidp}. From above,
    we know that rule~\ref{spoofed} is in conflict with rule~\ref{avoidp}, the latter is  stronger. Hence, the priority rule $C1$
    says that B4 is stronger than B2, $B4>B2$.

 \subsection{Further Reasoning Rules and Priorities}
 Argumentation reasoning is a high level approach that 
   makes it easier to get human specifications
and to modularly update the system when the specifications change.
 Its modularity  makes argumentation well suited for working with the attribution problem, where the evidence is incomplete and conflicting,
and also new evidence and reasoning rules can be introduced, together with new conclusions. 
It is important to note that attribution in cyber attacks will never be a final perfect theory, 
it will always be a heuristic
and evolving theory. 
Argumentation can take in new arguments without the need to reconsider the old argument representation and
preferences between them. One only needs to consider the preferences of the new arguments.

 We describe below further reasoning rules extracted from the analyses of 
 other cyber attacks, i.e., Stuxnet~\cite{stuxnet} and APT1~\cite{apt1}, and their context dependent priorities.
The different arguments on attribution have different strength/priorities. Below we list arguments, that conclude
 that a certain country performed the attack (except for argument 2 and 7, that conclude the opposite) by priority, from the highest to
the lowest, where the order of the rules introduced in the above subsection is not touched. 
\begin{enumerate}
\item \emph{Avoid(Attack, Country)} is a strong argument, e.g., the code was written to avoid  damage to that country. 
\item \emph{Spoofed(IP)} is stronger than the coming predicates
 because it shows that IP cannot be used as a credible evidence for the source of the attack.
\item \emph{SourceIP(Attack, IP)} combined with \emph{geoloc(IP, Country)} is a good argument\footnote{If the IP is not spoofed or anonymised.} as provides evidence for the attacker. 
\item \emph{Language(Attack, Country)} means that if the language of the code is the same as the language of the country, then that country performed the attack.
\item \emph{Motive(Country, Attack)} means that a country has the \emph{motivation} for performing the attack. 
\item \emph{Capable(Attack, Country)} means that a country is \emph{capable} of performing the attack (has the necessary skills, funds and resources/infrastructure). 
\item \emph{Target(Attack, Country)} is a weak argument. It says that if that particular country was the target of the attack, then it did not perform the attack.
\end{enumerate}

The above rules are given to GorgiasB, which will acommodate easily the new rules. 
Suppose we find evidence of an attack that mainly occurred in country $C$, a set of $IPs$ performed the attack, and
 good part of them are located in $C1$, and $C1$ is a rival of $C$. 
GorgiasB  will analyze the pieces of evidence by using the above arguments, and will state: $perform(Attack, C1)$.
Further evidence shows that part of the $IPs$ linked to $C1$ are spoofed. Hence, $C1$ did not performed the attack.
There is a conflict between arguments, which is solved by  GorgiasB through the use of the above priority rules. Therefor,
as spoofed is a stronger argument than the rest, the conclusion is $C1$ did not performed the attack.
Suppose now that traces of language $L2$ are found, where $L2$ is spoken in $C2$.
By re-checking the locations of the attacker $IPs$ a part of them are located in $C2$, and $C2$ has motives for attacking 
$C$. When the new evidence is given to GorgiasB, it concludes $perform(Attack, C2)$. 

\section{Data Sharing Based on Argumentation Decision Process}
The increase of connectivity and sharing of Data over 
 cloud environments brings positive impacts for 
 social good. 
Data has become dynamic and distributed, and  easily accessible from remote locations. 
Citizens in remote areas of the world
 are able to vote, follow courses, and work remotely.  
Another important use of dynamic and distributed data is in e-health, where patients can
 have an health care service, even if the first nearby
hospital is hundred of kilometres away.
For example, the doctor visiting the patients in remote areas, 
is able to access the patients medical data, perform remote visits to patients, access their 
data\footnote{Think about the vast technologies of self-test/analyses that the patients can perform.} and change their prescriptions 
depending on the self-tests/analyses the patients have made.

\subsection{The Problem of Data Sharing}
Data is shared as well as stored, used and transformed, not only by users
but also by companies, organisations and governments of different countries. There is a need of not just protecting the data,
but also for a correct and effective sharing of data by respecting the end user requirements, 
the different business rules (who are processing, using and sharing the data), and the legislation rules (the legal rules depending on the type of data,
the citizen rights, the business obligations and rights, the geographical location or origin of the data, where the data is shared or processed, etc.).
Being able to respect the above rules and requirements is not trivial, as dealing with  
 different legislations and domains can create conflicts between which rules to apply. For example, which rules of data sharing should be
 applied to the doctor that wants to access the patients data, that is hospitalized in an hospital, not in the patients residential country.

When an exchange of data occurs the two parties should agree on the rules related to the data, and the rules should be enforced. 
The first part is called the data sharing agreement (DSA)~\cite{Swarup06} where the different rules are specified and the two entities need to agree on it.  
 The rules express how and who is permitted/denied/obliged to access, delete, and share the data, as well as the different constraints that should be 
respected, e.g., time constraints, geographical ones. We will focus on data sharing techniques that use \emph{policies}
 for describing the different terms. 
The enforcement~\cite{dusagecontrol,usagecontrol1} is done when the data is used, 
and it identifies if the user/application is granted access to the data
and what level of access is allowed.

\paragraph{Data Sharing with Argumentation}
Non-monotonic reasoning, in particular argumentation, 
is broadly known for its use in representing legal rules and in constructing
decision processes where conflicting rules are involved. We propose a technique based on argumentation reasoning, that 
given the different policies that regulate how the data can be accessed/used/shared, finds the different conflicts between arguments
and creates a hierarchical decision process model, that decides which policy should be applied in any particular case. 
 This is the first attempt, to our knowledge, that uses argumentation reasoning for solving 
problems of data access and usage control.
The problem of data access and usage control is well studied~\cite{usagecontrol1,cnl4dsa}, 
but the existing solutions do not permit conflicting rules, together with a solution to the conflicts.

We represent the different policies that regulate the data sharing as argument rules. The preconditions
of the policy represent the premises of the argument rule, while the conclusion of the policy is the conclusion of the 
argument rule. Two policies are in conflict if they both permit and deny (or deny and oblige) to perform an action for certain preconditions,
that can be the same, or related between them. We use the GorgiasB tool for constructing the decision process.
\subsection{An Example of Decision Process for E-Health}
We show how we can represent different policies as arguments that are
given as input to GorgiasB, which
 finds the conflicting rules. 
The resolution of the conflicts is done by introducing priority rules and specifying 
the context when an argument is stronger than another one.  
 GorgiasB re-checks the arguments and the priority rules, if any conflict is detected,
then other priority rules are introduced.
 
Let's introduce an example of an e-health scenario. We have the patient, denoted by $C$, who is the owner of the
data, denoted by $Data$, and the relation  $Owner(C, Data)$. 
The data can be of three types: 
 (i) prescriptions: $Presc(Data)$, e.g., blood pressure, medicine prescriptions, x-rays;
(ii) private prescriptions: $PData(Data)$, e.g., anti-depressive treatments;
(iii) personal information: $PInfo(Data)$, e.g., address, family contacts. 
 The policies  give as conclusion the access predicate: $Access(A, B, permitted/denied)$,
where actor $B$ is $permitted/denied$ access to data $A$.

The data can be accessed by the doctor, denoted by $D$, that is a family doctor of the patient ($famD(D,C)$),
or a treating doctor ($treatD(D,C)$). Usually the family doctor has access to the patients' data,
while the treating doctor can access just to the prescriptions. This general part of the access
policies is represented by the following argument rules:
\begin{displaymath}
\renewcommand\arraystretch{0.9}
\begin{array}{rl}
Access(Data, D, permitted) \leftarrow famD(D, C) \ \wedge & \\ Owner(C, Data) \wedge Presc/PData/PInfo(Data) &   (1) \\
Access(Data, D, permitted) \leftarrow treatD(D, C) \ \wedge & \\ Owner(C, Data) \wedge Presc(Data) & (2)\\
Access(Data, D, denied) \leftarrow treatD(D, C) \ \wedge & \\ Owner(C, Data) \wedge PData(Data)  & (3)\\
Access(Data, D, denied) \leftarrow treatD(D, C) \ \wedge & \\ Owner(C, Data) \wedge PInfo(Data)  & (4)
\end{array}
\end{displaymath}
If the patient is an emergency situation ($Emerg(C)$), then the treating doctor can access the private
 prescriptions and information
   e.g., for contacting a patients family member.
\begin{displaymath}
\renewcommand\arraystretch{0.9}
\begin{array}{rl}
Access(Data, D, permitted) \leftarrow Emerg(C) \ \wedge & \\  treatD(D, C) \wedge Owner(C, Data) \wedge PData(Data) &   (5)\\
Access(Data, D, permitted) \leftarrow Emerg(C) \ \wedge & \\  treatD(D, C) \wedge Owner(C, Data) \wedge PInfo(Data) & (6)
\end{array}
\end{displaymath}
When we give the above rules to GorgiasB, it finds two conflicts.
Rules (3) and (5), and rules (4) and (6) are in conflict,
 as one says that generally treating doctors cannot access the private prescriptions/information of a patient, 
while the second one gives permission to the treating doctor to access them in case of an emergency. 
 Two priority rules are introduced: in case of an emergency rule (5) is preferred over rule (3), (rule(5) $>$ rule(3)), 
 and rule (6) is preferred over rule (4), (rule(6) $>$ rule(4)). 
 After the preferences are introduced, GorgiasB does not find any conflict. 
 
 For obvious reasons, a patient can access to her/his own data.
 In European countries when the patient is suffering from depression, 
 or its medical data can be of high emotional impact 
 $Emot(C)$ (e.g., suspects of a
 terminal illness that if revealed to the patient early can effect the wellbeing and the patients life), 
 then s/he is not allowed to access the data.  
 \vspace{-2pt}
\begin{displaymath}
\renewcommand\arraystretch{0.9}
\begin{array}{rl}
Access(Data, C, permitted) \leftarrow Owner(C,Data) \wedge & \\ Presc/PData/PInfo(Data)& (7)\\
Access(Data, C, denied) \leftarrow Emot(C) \ \wedge & \\ PData(Data) \wedge   Owner(C, Data)  & (8)
\end{array}
\end{displaymath}
In this case, rules (7) and (8) are in conflict between each other, and rule (8) $>$ rule (7),
in the context of $Emot(C)$. 

 When a patient is hospitalized,
s/he agrees on a set rules regarding the intensive care.
In case the patient is in intensive care, $Intens(C)$, the treating doctor can access the patents prescriptions, 
but not the rest of the data (compliant with rules (2-4)).
\vspace{-2pt}
 \begin{displaymath}
 \renewcommand\arraystretch{0.9}
\begin{array}{rl}
 Access(Data, D, permitted) \leftarrow Intens(C) \ \wedge & \\ treatD(D, C) \wedge Owner(C, Data)\  \wedge & \\  Presc(Data)& (9)\\
  Access(Data, D, denied) \leftarrow Intens(C)\  \wedge & \\ treatD(D, C) \wedge Owner(C, Data)\ \wedge & \\  PData/ PInfo(Data) & (10)
\end{array}
\end{displaymath}
In special cases of intensive care, when further diagnosis needs to be conducted,
the treating doctor may  need to access also to the 
private prescription of the patient.  For doing that, he needs the patient permissions, $Perm(C, PData)$.
\vspace{-2pt}
 \begin{displaymath}
  \renewcommand\arraystretch{0.9}
\begin{array}{rl}
 Access(Data, D, permitted) \leftarrow Intens(C)\  \wedge & \\ Perm(C, PData) \wedge treatD(D, C) \ \wedge & \\ 
 Owner(C, Data) \wedge PData(Data) & (11)
\end{array}
\end{displaymath}
Rules (11) and (10) are in conflict, as the first is permitting the access while the second is denying it. 
Rule (11) is more specific, thus, rule (11) $>$ rule (10).

Suppose that the patient is unconscious, or unable to give the permission for the private prescription,  $Uncon(C)$. 
In this case, the treating doctor has the right to access the patient personal information, 
and contact a family member and get
the permission from them ($fPerm(C, PData)$).
\vspace{-4pt}
 \begin{displaymath}
  \renewcommand\arraystretch{0.9}
\begin{array}{rl}
 Access(Data, D, permitted) \leftarrow Intens(C)\ \wedge & \\
 Uncon(C) \wedge  treatD(D, C)\ \wedge & \\  Owner(C, Data) \wedge PInfo(Data)& (12)\\
  Access(Data, D, permitted) \leftarrow Intens(C)\ \wedge\\ 
  Uncon(C) \wedge fPerm(C, PData) \ \wedge & \\  treatD(D, C) \wedge Owner(C, Data) \ \wedge & \\ PData(Data) & (13)\\
\end{array}
\end{displaymath}
Rules (12) and (10) are in conflict: the first permits the access to the private information while the second one denies it.
In this case, rule (12) $>$ rule (10). Also, rules (13) and (10) are in conflict: similarly rule (13) $>$ rule (10).

Suppose the family member cannot give the needed permission.
In this case, the hospital regulation says that 
a double permission by the family doctor ($fdocP(C, PData)$) 
and the head of the hospital ($hdP(C, PData)$) is needed. 
\vspace{-4pt}
 \begin{displaymath}
  \renewcommand\arraystretch{0.9}
\begin{array}{rl}
  Access(Data, D, permitted) \leftarrow Intens(C) \ \wedge & \\
   Uncon(C) \wedge  \neg fPerm(C, PData)\  \wedge & \\
       fdocP(C, PData)  \wedge  hdP(C, PData) \ \wedge  &  \\
   treatD(D, C)  \wedge  Owner(C, Data) \ \wedge & \\  PData(Data) & (14)
\end{array}
\end{displaymath}
Rules (14) and (10) are in conflict, as the first permits access to the private prescription while the second denies it. 
In this case, rule (14) $>$ rule (10), as rule (14) is more specific.

GorgiasB detects all the above conflicts, takes in new rules,
analyze them and introduces priority rules.  For the above cases, the specificity between the premisses of the arguments,
has been the winning factor between rules. 

  \vspace{-2pt}
\section{Future Work and Conclusion}
There are several challenges to be addressed in our approach so that we can use
argumentation as a basis for tackling security and privacy problems
in the new highly interconnected cyber world. The two specific
problems that we have presented above can guide us in developing a
general methodology for argumentation-based security applications.
We aim to study such general methodologies and use them 
to analyze and solve other problems such as the distributed firewall problem,
cyber attack detection, protective measure against attacks,
disaster management response. 

The two particular applications of cyber attack attribution and 
regulatory data sharing allow us 
to develop a good understanding of the general challenges that 
we face. The attribution problem is based on 
knowledge on the behaviour of attacks that is coarse and to
a certain degree speculative, e.g. knowledge that comes from 
analytics that indicates frequently occurring patterns but by
no means strict and universal mode of attack behaviour. In the
case of data sharing the knowledge is quite well documented
in regulation and/or legislative policies but can have quite 
a complex form especially when several policies need to 
be considered simultaneously. 
In both areas there are different problems that 
can be considered for improving our approach.

\textbf{Future work of argumentation  in attribution}
Our cyber attacks attribution reasoner uses the available evidence for
reaching conclusions about who did the attack. The argumentation
approach provides an explanation to justify the various
conclusions reached. We are working on extending the reasoner for
guiding the analysts during the forensics investigation and
advising where and what to look for. An automatic collection
and analysis of the evidence passed to the reasoner
would increase its efficiency.
We plan  on further extending our work for representing
quantitatively the strength of arguments. Thus, to
quantitatively compare the arguments for different contexts,
and define a more granular priority between argument rules.
This would allow us to be more precise in our assessments,
than a binary decision e.g. of simply whether a country had performed
the attack or not.

The reasoning rules that were extracted by
real examples should be extended. This can be done by analyzing
other examples and by inferring the used rules. An
important task for future work would be to consider evidence
of symptoms of attacks.  In this way, the reasoning process, if applied at run-time,
can help in detecting ongoing attacks and can guide the analyst during the process
of evidence collection.
We plan on working on human cognitive
reasoning for capturing the evidence of the
strategic layer, called also social evidence.

\textbf{Future work of argumentation in Data sharing}
With our decision process model based on argumentation reasoning
we find the conflicting policies and introduce the priority
rules. 
 The policies used
by this decision process deal with the data access
and usage control. It would be useful to work with
policies that deal with data integrity and availability.

For now the conflict resolution is made manually, by specifying
contextual priorities between argument rules. A future challenge is to
make the process fully automated, by introducing general
rules for conflict resolutions that will permit the automatic
selection, e.g., a generic rule that was used for the above examples
was the specificity of the premisses used by the arguments.
Another future challenge is the use of online learning,
where depending on the context, we can
learn which are the rules that have higher priority.

\textbf{Conclusion}
We presented two important problems for social good: cyber attack attribution
and regulatory data sharing. Both problems can be seen as decision making problems:
what rules and evidence to use for detecting the attacker; which rule
to use while using and sharing data. 
One of the main challenges that these problems pose is 
the used information that can be incomplete, conflicting and context dependent.
The proposed solution is based on
argumentation reasoning, 
that is an appropriate technique for implementing decision making mechanisms 
under incomplete and conflicting information.
Our solution solves the problems of  attribution in cyber attacks
and regulatory data sharing. 

   \vspace{-2pt}
\section*{Acknowledgments}
Supported by FP7 EU-funded project Coco Cloud grant no.610853, and EPSRC Project CIPART grant no.~EP/L022729/1.

\bibliographystyle{aaai}
\bibliography{arg}

\end{document}